\documentclass[prl,twocolumn,floatfix, footinbib, article, showpacs]{revtex4-1}

\newcommand{\be}{\begin{eqnarray}}
\newcommand{\ee}{\end{eqnarray}}
\newcommand{\ket}[1]{\ensuremath{\left| {#1} \right>}}
\newcommand{\bra}[1]{\ensuremath{\left< {#1} \right|}}
\newcommand{\braket}[2]{\ensuremath{\left< \left. {#1} \right| {#2} \right>}}

\usepackage{graphicx}
\usepackage{amsmath}
\usepackage{amssymb}

\begin{document}

\title{Experimental recovery of a qubit from partial collapse}

\author{J.A. Sherman}
\email{Email: jeff.sherman@nist.gov}
\author{M.J. Curtis}
\author{D.J. Szwer}
\author{D.T.C. Allcock}
\author{G. Imreh}
\author{D.M. Lucas}
\author{A. M. Steane}
\affiliation{Clarendon Laboratory, Department of Physics, University of Oxford, Parks Road, Oxford OX1 3PU, UK}

\begin{abstract}
We describe and implement a method to restore the state of a single qubit, in principle perfectly, after it has partially collapsed. The method resembles the classical Hahn {\em spin-echo}, but works on a wider class of relaxation processes, in which the quantum state partially leaves the computational Hilbert space.  It is not guaranteed to work every time, but successful outcomes are heralded.  We demonstrate using a single trapped ion better performance from this recovery method than can be obtained employing projection and post-selection alone.  The demonstration features a novel qubit implementation that permits both partial collapse and coherent manipulations with high fidelity.
\end{abstract}

\date{\today}

\maketitle

The concept of {\em spin-echo}~\cite{hahn1950spin,meiboom1958modified} is, at heart, the fact that
\be 
\left(e^{i \phi \sigma_z} \sigma_y \right)^2 = I \quad (\text{for all }\phi),          \label{spinecho}
\ee
and the observation that Eq.~(\ref{spinecho}) has important practical applications~\cite{derome1987modern}. Here, $I$ is the identity operator and $\sigma_{y,z}$ are Pauli matrices. The first term, $\exp(i \phi \sigma_z)$, might be a spin rotation caused by an uncontrolled magnetic field whose size is unknown or which varies across a sample, but which is constant (or almost constant) in time.  The second term, $\sigma_y$, represents a 180$^\circ$~rotation about $\hat{y}$ and results from the experimenter applying a $\pi$-pulse to the spin system. The square indicates that the uncontrolled interaction influences the system again for the same duration (or, more generally, integrated strength), thus introducing a further uncontrolled phase. However, thanks to the $\pi$-pulse, the second phase `unwinds' the effect of the first. A final $\pi$-pulse (included here, optional in practice) produces a simple overall outcome, the identity operation. In the language of error correction, spin-echo is a way of recovering from one type of correlated (i.e.\ same during the two parts of the echo), unitary error process.

Here we consider a certain non-unitary error process and a recovery method that can be understood either as a generalization of spin-echo, or as a form of quantum error detection, or as an example of weak measurement~\cite{ritchie1991realization}, or as an `uncollapse' process, to use terminology adopted by Katz {\em et al.}~\cite{katz2006coherent,katz2008reversal}. The process is not originated by us;  it is described in~\cite{wu2002efficient,byrd2005universal} and pursued in~\cite{katz2008reversal}. We discuss the theory more generally, and present a more complete and accurate experimental realization, reducing the infidelity of the process by an order-of-magnitude. In contrast to other recovery schemes~(e.g.\ \cite{keane2012simplified,schindler2013undoing}), only one physical qubit is required. Importantly, our data show that the recovery process fidelity exceeds that which is obtained by filtering with projection alone. 

Consider now a two-state system in which one state is unstable. We can model this as a three-state system in which the first two states, $\ket{0}$, $\ket{1}$ form the computational Hilbert space ${\cal H}_L$, and the third state, $\ket{2}$, accounts for unspecified further degrees of freedom (dimensions in Hilbert space). The error process we have in mind is incoherent population transfer from $\ket{1}$ to $\ket{2}$ with probability $p$; the effect on the system's density matrix $\rho$ is
\be
{\cal C}_p \left( \rho \right) = \left( \begin{array}{ccc}
\rho_{00} & \rho_{01} \sqrt{1-p} & \rho_{02} \\
\rho_{10}\sqrt{1-p} & \rho_{11}(1-p) & \rho_{12} \sqrt{1-p}  \\
\rho_{20} & \rho_{21} \sqrt{1-p} & \rho_{22} + p \rho_{11}
\end{array} \right).
\ee
Note, we do not require that the system have exactly three states, only that this mapping correctly characterizes the loss of population from ${\cal H}_L$.

If the system's initial state is in ${\cal H}_L$, then such a process may be regarded as a non-unitary `leakage error'~\cite{fazio1999fidelity} occurring with probability of order $p$.  In a computational setting, one can manipulate the qubit but not the environment.  So it is possible, for example, to make a measurement such that the state is projected onto either ${\cal H}_L$ or the orthogonal space.  If the state was initially prepared in ${\cal H}_L$, then after such a measurement, in those cases where the final state is projected back onto ${\cal H}_L$ (an outcome with probability $1- \rho_{11}p$), the net effect is a transformation of the qubit's density matrix:
\be
 {\cal M}_p \left(\rho \right) = \frac{1}{1 - \rho_{11} p}
\left( \begin{array}{ccc}
\rho_{00} & \rho_{01} \sqrt{1-p} \\
\rho_{10}\sqrt{1-p} & \rho_{11}(1-p) 
\end{array} \right).
\ee
When $p=1$, $ {\cal M}_p$ projectively measures (`collapses') the qubit's state; when $p<1$, we may call the process a \emph{partial measurement} or \emph{partial collapse}. Pure initial states remain pure afterwards, but the ${\cal M}_p$ process is non-linear and can be seen as a non-unitary movement of the Bloch vector towards the $\ket{0}$ direction.  The restriction to cases where the final state remains in ${\cal H}_L$ is a \emph{post-selection}. 

%


\begin{figure}
\includegraphics[width = \linewidth]{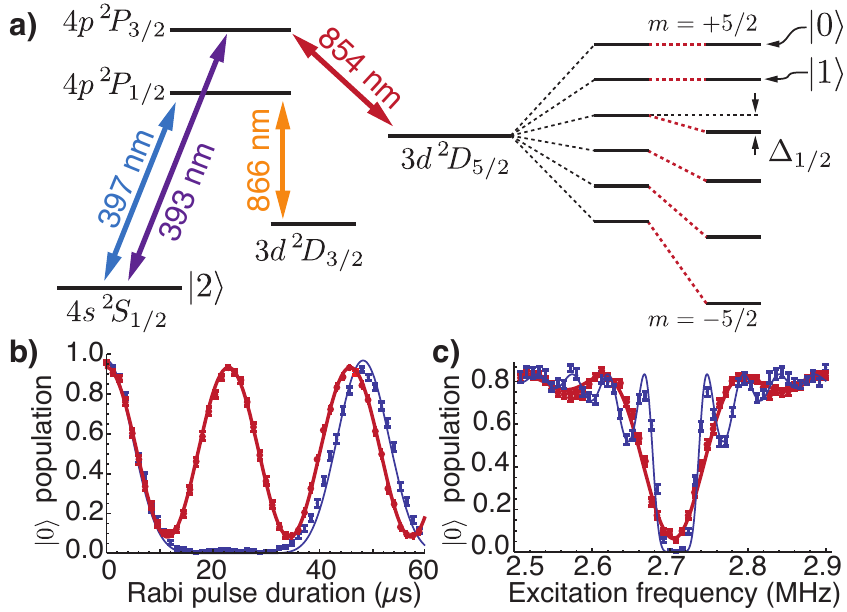}
\caption{(color online) (a) Relevant levels and transitions in $^{40}$Ca$^+$. The Zeeman substructure of $3d\,^2\!D_{5/2}$ is shown without and then with the light-shifts $\Delta_m$ introduced by an intense, circularly-polarized 854~nm laser. Blue data (and thin lines) show spin-precession (Rabi flopping) as a function of (b) time and (c) excitation frequency in the unshifted $D_{5/2}$ manifold.  Red data (and thick lines) demonstrate the isolation of the two-state qubit due to the applied light-shift (ac-Stark effect).  The lines are numerical fits to Bloch equations.  As expected, the isolated qubit spin-precession rate exceeds that of the unshifted $D_{5/2}$ manifold by a factor of $\sqrt{5}$~\cite{rose1957elementary}.}
\label{fig:levels}
\end{figure}

The recovery process we discuss and implement here is
\be
 {\cal R}_p \left( \rho \right) &=& \sigma_y  {\cal M}_p \left( \sigma_y  {\cal M}_p(\rho) \sigma_y^\dagger \right) \label{eq:R}
\sigma_y^\dagger, \\
&=& \rho  \;\; \mbox{when} \;\; p < 1.
\ee
We let the leakage error act twice, sandwiched between  $\pi$-pulses and projections, the overall effect of which is the identity operation. 

A useful notation is provided by writing
$
 {\cal M}_p \left(\rho \right) = M \rho M^\dagger / N_1,
$
where $N_1 = {\rm Tr} (M \rho M^\dagger) = 1- \rho_{11} p$ is a normalization factor, and $M = \ket{0}\bra{0} + \sqrt{1-p} \ket{1}\bra{1}$ is a {\em measurement-} \cite{nielsen2010quantum} or Kraus-operator~\cite{kraus1971general}.
$M$ describes the relaxation process (the modeled error) combined with the projection into ${\cal H}_L$ (part of the recovery protocol). The complete recovery protocol ${\cal R}_p$ works because
\be
(N_1 N_2)^{-1/2} \, \left( M e^{i \phi \sigma_z} \sigma_y \right)^2 = I,
\ee
where $N_2 = 1 - \rho_{00} p / N_1$. Like spin-echo, ${\cal R}_p$ relies on the uncontrolled error process being the same in two successive time intervals. The method may prove to be useful in practice because it retains a large degree of generality. It is independent of the qubit state and the conditional outcome is independent of the strength of the error process (i.e. $p$)---the only thing that depends on $p$ is the probability of obtaining the desired outcome, which is $1-p$ (regardless of input state).

As noted, Katz {\em et al.}\ referred to  ${\cal R}_p$ as an `uncollapse'. We prefer the terminology `filtering' or `error detection'.  A quantum error detection is, in general, a process in which a device interacts with a communication channel in such a way as to signal when the channel introduces errors, without (as far as possible) corrupting those cases where no error occurs. A simple means of error detection in the present case consists of performing a projective measurement after one use of the channel (i.e.\ one occurrence of ${\cal C}_p$) and accepting the resulting state if it was projected onto ${\cal H}_L$. The outcome of this strategy is given by ${\cal M}_p (\rho)$: when the projection is successful (i.e.\ the detector reports `no error'), one obtains the final state $\ket{d} = (1 - |b|^2 p)^{-1/2} \left( a \ket{0} + b \sqrt{1-p} \ket{1} \right)$ given an initial state $\ket{i} = a \ket{0} + b \ket{1}$. The fidelity of such an accepted state is
\be
F_  {\cal M} = |\braket{i}{d}| = \frac{|a|^2 + |b|^2 \sqrt{1-p}}{\sqrt{1-|b|^2 p}}. \label{fid} 
\ee
Since the infidelity $1-F_  {\cal M} = |a|^2 |b|^2 p^2 / 8 + O(p^3)$  is of order $p^2$, the process ${\cal M}_p$ is said to be `single-error-detecting'. The advantage of the process ${\cal R}_p$ over ${\cal M}_p$ is that the final state infidelity is strictly zero when the detector reports `no error'. Such a filter makes overall use of the channel noise-free in cases where it succeeds.

It is interesting to ask whether ${\cal R}_p$ [Eq.~(\ref{eq:R})] can achieve a higher fidelity, even with experimental imperfections, than would be possible using the single projection and post-selection ${\cal M}_p$ [Eq.\ (\ref{fid})].  Whereas a previous implementation of ${\cal R}_p$~\cite{katz2008reversal} did not demonstrate this, here we show that it can.

We used a single trapped and laser-cooled $^{40}$Ca$^+$ ion in our experiments~\cite{curtis2010thesis}. The main issue for the physical implementation was the need for a near-ideal projective measurement: one which could both project the state onto (or perpendicular to) ${\cal H}_L$ and also give a large enough detectable signal so that we know the measurement outcome, preferably in a single shot, without disturbing the state {\em within} ${\cal H}_L$.  Optical pumping can project states in single atoms, but this does not in itself guarantee a detectable signal since the number of scattered photons may be too low. For this reason, for example, we could not easily employ the two spin states of the ion's ground state as our qubit, as was done previously~\cite{mcdonnell2004high-efficiency}. Instead, we adopted two Zeeman sublevels $m=3/2$ and $5/2$ of the metastable $3D_{5/2}$ level of $^{40}$Ca$^+$. 

We manipulate the qubit (e.g.\ provide $\pi$-pulses) by driving spin transitions with a radio-frequency (RF) coil~\footnote{The 10-turn RF coil (70~mm diameter), was $\approx$~50~mm from the ion. It was connected to an air-gap variable capacitor adjustable over 10--225 pF. This resonant circuit tuned between 2.5 and 6 MHz, with a quality factor of order 10. A pair of phase-locked signal generators provided waveforms for the pulse sequences. We adjusted the pulse amplitude to obtain a Rabi frequency of order 50 kHz.}. However, at low magnetic field (0.16 mT), all the magnetic-dipole transitions within $D_{5/2}$ are resonant with the qubit splitting $\omega / 2 \pi \approx 2.7$~MHz. To isolate the qubit from the states $m < 3/2$, we applied an intense circularly-polarized $(\sigma^+)$ laser detuned from the $D_{5/2}$--$P_{3/2}$ transition wavelength (854~nm). As Figure~\ref{fig:levels} shows, this introduces light-shifts $\Delta_m$ for levels $m < 3/2$, but not the qubit levels owing to angular momentum selection rules~\cite{delsart1980effects,deng2001light,stalnaker2006dynamic}. With a beam power of 15 mW (an intensity $\sim 1.5 \times 10^7$~W/m$^2$), and detuning $\approx$~100 GHz, we obtained a qubit-isolating light-shift $\Delta_{1/2}/2 \pi \approx 1$ MHz.

The advantage of our chosen qubit is the ease with which it is measured, and, in particular, the ease with which partial measurements are performed.  We label qubit states $\ket{0}, \ket{1} \equiv \ket{D_{5/2}, m=5/2}, \ket{D_{5/2}, m=3/2}$. A measurement proceeds by first applying a `deshelving' pulse: a weak laser pulse at 854~nm, polarized mainly $\pi$ with some $\sigma^+$ and resonant with $D_{5/2}$--$P_{3/2}$. It couples to $\ket{1}$ but not to $\ket{0}$, owing to angular momentum selection rules, and causes optical pumping to the ground state, $4S_{1/2}$. We then detect whether optical pumping occurred by driving $4S_{1/2}$--$4P_{1/2}$ (397~nm `cooling') and $3D_{3/2}$--$4P_{1/2}$ (866~nm `repumping') transitions with a pair of lasers while collecting 397~nm fluorescence photons. An atom in $4S_{1/2}$ scatters many photons, yet the fluorescence does not disturb the qubit states, so the fluorescence detection stage is, to good approximation, perfect~\cite{myerson2008high,keselman2011high}.

In contrast, the initial deshelving is sub-optimal due to the non-zero probability for decay from $P_{3/2}$ to $D_{5/2}$. Decay from $P_{3/2}, m' = 3/2$ to $D_{5/2}, m \le 3/2$ is harmless; it only lengthens slightly the optical pumping time. However, a rare decay from $P_{3/2}$ to $\ket{0}$ is problematic. The result is that instead of ${\cal C}_p$, we experimentally realize
\[
{\cal D}_p \left( \rho \right) = \left( \begin{array}{ccc}
\rho_{00} + \epsilon p \rho_{11} & \rho_{01} \sqrt{1-p} & \rho_{02} \\
\rho_{10}\sqrt{1-p} & \rho_{11}(1-p) & \rho_{12} \sqrt{1-p}  \\
\rho_{20} & \rho_{21} \sqrt{1-p} & \rho_{22} + (1-\epsilon) p \rho_{11}
\end{array} \right),
\]
where $\epsilon = 0.0355$ is the unfavorable branching ratio. Thus we do not implement the ideal process ${\cal R}_p$ but a related process ${\cal R}'_p$ which approximates it if $\epsilon p \ll 1-p$.

The high reliability of the fluorescence measurement aided in optimizing pulse parameters and alignment. Viewport birefringence complicates specifying the polarization of laser beams at the ion.  However, optical pumping experiments~\cite{curtis2010thesis} enabled us to trim the de-shelving laser's polarization impurity to $5 \times 10^{-4}$ in intensity.

\begin{figure}
\includegraphics[width= \linewidth]{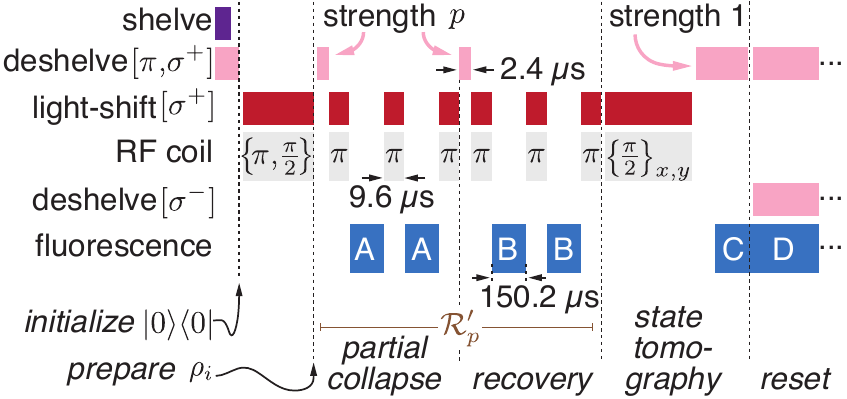}
\caption{(color online) Experimental sequence for testing ${\cal R}'_p$. The `shelve' laser is at 393$\,$nm; the others are as described in the text.  Post-selection filters away attempts when 397~nm fluorescence is found in detection intervals $A$ or $B$.  Scattering during $D$ re-cools the ion to near the Doppler-limit.}
\label{f.seq}
\end{figure}

\begin{figure*}
\includegraphics[width = \linewidth]{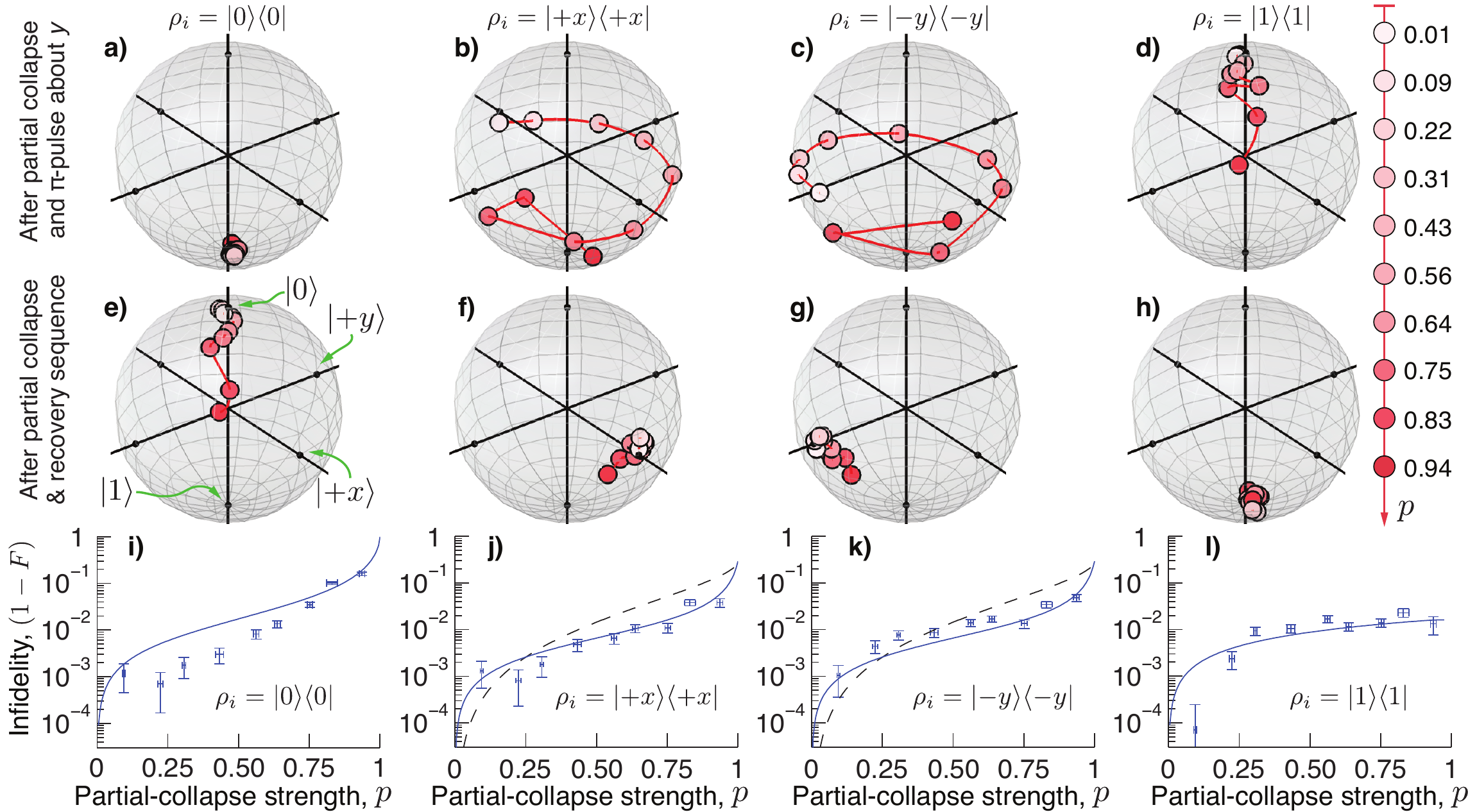}
\caption{(color online) Observed states on the Bloch sphere after `partial-collapse' (a-d), including an odd number of $\pi$-pulses (see Figure~\ref{f.seq}), and after `recovery' (e-h) for four initial states $\rho_i$. Points are tomography results for 10 partial measurement strengths $p \in (0.01,\ldots,0.94)$; lines guide the eye. The evident spiraling is an azimuthal phase that can be compensated by spin-echo; the increasing polar angle seen in (b-d) is the `partial-collapse' process. After recovery, most data clusters around the correct input state.  $\ket{0}$ at $p > 0.7$ performs less well owing to the finite branching ratio $\epsilon$ in ${\cal D}_p$; in this case $\rho_f$ moves down the Bloch sphere's axis, reaching its center at $p\simeq 0.97$. (i-l) Measured recovery process infidelity; vertical and horizontal error bars indicate the shot noise and uncertainty in $p$, respectively.  Full lines are predictions of ${\cal R}'_p$ with no free parameters.  Dashed lines depict the infidelity of a simpler projection and post-selection, ${\cal M}_p$.  Note: for states $\ket{0}$ and $\ket{1}$, $1 - F_{\cal M} = 0$ is off-scale.}
\label{f.bloch}
\end{figure*}

Figure~\ref{f.seq} shows the experimental sequence used to implement ${\cal R}'_p$. We prepared initial states by first optically pumping to $\ket{0}$, then applying a $\pi/2$- or $\pi$-pulse when necessary to the RF coil.  To help protect the qubit from magnetic field fluctuations during the measurement periods, we employed a sequence of three further $\pi$-pulses to provide dynamical decoupling (in the CPMG~\cite{freeman2003spin} timing). We tested a set of four different initial states and employed quantum process tomography~\cite{nielsen2010quantum} to fully characterize state preparation, the partial collapse (without recovery), and the full process ${\cal R}'_p$.

Figures \ref{f.bloch}(a-d, e-h) illustrates the Bloch vector just after a $p$-strength partial collapse, and after recovery, as deduced by tomography. At the intermediate stage the state has moved towards one pole of the Bloch sphere, owing to the relaxation followed by projection; after `recovery' the Bloch vector is back near its initial position.

Figures \ref{f.bloch}(i-l) plot the measured infidelity $1-F$ of the final states $\rho_f$  compared to initial states $\rho_i$, defined by
\be
F \equiv \mbox{trace} \left[ \left( \sqrt{\rho_i} \rho_f \sqrt{\rho_i} \right)^{1/2} \right].
\ee
We deduced initial states $\rho_i$ from the tomography output with $p=0$ (i.e.\ the deshelving pulse was not applied).  Therefore, $F = 1$ at $p=0$ by definition.  This choice allows us to extract the infidelity of the partial collapse and recovery process itself, independent of systematic state-preparation errors.  If instead we compare measured final states with the nominally intended initial states, we observe a further 1\% reduction in fidelity unrelated to $ {\cal R}'_p$.  Each data point is the average result of around 5,000 repetitions of the experimental cycle at $p=0.1$, rising to 12,000 at $p=0.9$.  Solid lines show the expected results for the process $ {\cal R}'_p$ (that is, the one implemented, in which error ${\cal D}_p$ not ${\cal C}_p$ occurs). These are not fits but predictions for a perfect implementation of $ {\cal R}'_p$, with no free parameters. The observed behavior matches the expectation well.

We observe fidelities close to unity even for relatively large partial-collapses $p > 0.5$, suggesting we achieve substantial recovery of the qubit's state. However, to assess this claim more carefully, one should ask whether the `recovery' step has in fact made matters better or worse. A suitable criterion is given by the dashed lines. These show the infidelity of the simpler strategy of merely projecting into ${\cal H}_L$ after a single interval of relaxation (possibly including a classical spin-echo), $1-F_{\cal{M}}$ [see Eq.~(\ref{fid})]. Arguably, only by exceeding this fidelity can one demonstrate any active recovery from the relaxation process, over and above that which is obtained by post-selection alone. Our experiment comfortably achieves this for states $\ket{x}$, $\ket{y}$, which are equally weighted superpositions of $\ket{0}$ and $\ket{1}$. For initial states $\ket{0}$ and $\ket{1}$, $F_{\cal M}=1$ so the `recovery' step can only make matters worse. However, in situations such as quantum computing, one would not know the qubit's initial state. A reasonable figure of merit is the average fidelity obtained for a set of states uniformly distributed over the Bloch sphere. For example, at a high partial collapse $p=0.8$, one finds for this average $\bar{F}_{\cal M} = 0.956$ and $\bar{F}_{\cal R'} = 0.986$.  After subtracting the predicted $F_{\cal R'}$ from our data, the residuals exhibit a standard deviation of $0.0097$. Thus the average fidelity inferred from the data exceeds $\bar{F}_{\cal M}$ by three standard deviations at $p = 0.8$, and systematically exceeds it over a range of $p$ values.  We stress that $\bar{F}_{\cal R'} \to 1$ in systems where the branching ratio $\epsilon \to 0$.

We conclude with some further general remarks. Suppose the relaxation process is exponential decay, such that $p = 1 - \exp(-\Gamma t)$, and we apply the recovery method $n$ times during a time interval $t$, in an ideal experiment. Then as $n \rightarrow \infty$  the probability of success tends to $e^{-\Gamma t/2}$. Thus one may interpret the process as one which makes symmetrical an otherwise asymmetric relaxation, such that population leaks equally out of the whole state space ${\cal H}_L$, and a projection back into that space suffices to restore the state. An alternative way to protect quantum information against the error process ${\cal C}_p$ is to encode a single logical qubit in a pair of physical qubits, using the states $\ket{0}\ket{1} \pm \ket{1}\ket{0}$. This is like a decoherence-free subspace~\cite{lidar1998decoherence}, but the amount of population remaining in the protected space decays with time.

We thank Nick Thomas for work on the vacuum apparatus and Derek Stacey for useful discussions.  This work was supported by the European Commission (SCALA), IARPA (ref.\ 47968-PH-QC), and EPSRC (QIP IRC).
\bibliography{errdetect2}

\end{document}